\documentclass[aps,prb,twocolumn,showpacs]{revtex4}
\usepackage{graphicx}
\begin{document}

\title{Kondo effect of Co adatoms on Ag monolayers on noble metal surfaces}

\author{M.A. Schneider, P. Wahl, L. Diekh\"oner, L. Vitali, G. Wittich, and K. Kern}
\affiliation{Max-Planck-Institut f\"ur Festk\"orperforschung,
Heisenbergstr. 1, D-70569 Stuttgart, Germany}

\date{\today}

\begin{abstract}
The Kondo temperature $T_K$ of single Co adatoms on monolayers of
Ag on Cu and Au(111) is determined using Scanning Tunneling
Spectroscopy. $T_K$ of Co on a single monolayer of Ag on either
substrate is essentially the same as that of Co on a homogenous
Ag(111) crystal. This gives strong evidence that the interaction
of surface Kondo impurities with the substrate is very local in
nature. By comparing $T_K$ found for Co on Cu, Ag, and Au
(111)-surfaces we show that the energy scale of the many-electron
Kondo state is insensitive to the properties of surface states and
to the energetic position of the projected bulk band edges.

\end{abstract}

\pacs{72.15.Qm, 73.20.Hb, 68.37.Ef}

\maketitle

The interaction of a magnetic impurity with the electrons of a
non-magnetic host serves as a paradigm of many-body physics. The
explanation of the resistivity minimum of dilute magnetic alloys
at low temperatures and the development of appropriate theoretical
tools to understand the rich phenomenology of the Kondo effect
\cite{Kondo, Anderson} are still shaping the research now done in
heavy fermion systems and Hi-Tc superconductivity.\cite{Hewson}
Even in the domain of the classical Kondo effect interest has been
renewed when two different experimental approaches became
possible: one was to study the Kondo effect in quantum dots which
behave like artificial "spin
impurities",\cite{G_G_QDot,Cronen_QDot,Weis_QDot} the other was
the use of the Scanning Tunneling Micrsoscope (STM) to study the
electronic properties of single magnetic atoms on metal surfaces.
\cite{Madhavan98,Li98} We present here results of the latter
technique.

The Kondo effect is due to the formation of a correlated singlet
ground state at low temperatures in which the magnetic spin of the
impurity is screened by a cloud of conduction band electrons interacting with the
impurity. The formation of this state lowers the energy of the
electronic system by the amount $k_BT_K$, where $T_K$ is the Kondo
temperature of the system which ranges from sub-Kelvin to several
hundred Kelvin. As function of temperature $T$ transport
coefficients like resistivity, specific heat, and susceptibility
depend on $T/T_K$, making $T_K$ the fundamental parameter of the
low energy excitations of the many-electron problem. The special
properties of the transport coefficients are caused by the
formation of a resonance at the Fermi energy in the single
particle density of states of the impurity. Its half width is
$k_BT_K$ for $T << T_K$.\cite{Hewson} Whereas transport
measurements probe this resonance indirectly it has been directly
probed by (normal and inverse) photoemission spectroscopy in the
case of Kondo alloys that contain rare earth metals
\cite{Reinert01} and by Scanning Tunneling Spectroscopy (STS) for
3d impurities at surfaces.\cite{Madhavan98,Manoharan00,
Jamneala00, Knorr02, Schneider02} It has been a first concern to
understand the mechanism by which the Kondo
resonance peak in the impurity density of states is transformed into
the Fano line shape seen by STS.\cite{Schneider02,Ujsaghy00,
Plihal01} In contrast, here we are concerned with understanding
what determines the Kondo temperature of a magnetic surface
impurity. Recently, we have put forward a simple model based on a
tight-binding approach to adatom hybridization that allows to
understand the trends in $T_K$ for Co on a variety of noble metal
surfaces.\cite{Wahl04} The experiments presented here support the
tight-binding view since it appears that in an overlayer system
$T_K$ is determined by the chemical identity of the first surface layer.

Single crystal surfaces were prepared by standard sputtering and
annealing cycles in ultra-high vacuum (base pressure $1\cdot
10^{-10}\mathrm{mbar}$). For the experiments on overlayer systems
Ag was evaporated from an electron beam heated source at room
temperature. The samples were then transferred \textsl{in situ} to
an STM working at $6~\mathrm{K}$. Co adatoms were produced by
dosing Co from a carefully out-gassed tungsten wire with a Co wire
of 99.99\% purity wrapped around it. During that process the
sample temperature stayed below $20~\mathrm{K}$ ensuring the
deposition of single adatoms due to a repulsive interaction between
them on noble metal (111) surfaces.\cite{Knorr02b} Spectroscopic
measurements were performed using a lock-in technique with a
modulation of the sample voltage of $1\mathrm{mV_{RMS}}$ at a
frequency of $4.5 \mathrm{kHz}$. All bias voltages are sample
potentials measured with respect to the tip.

\begin{figure}
\includegraphics{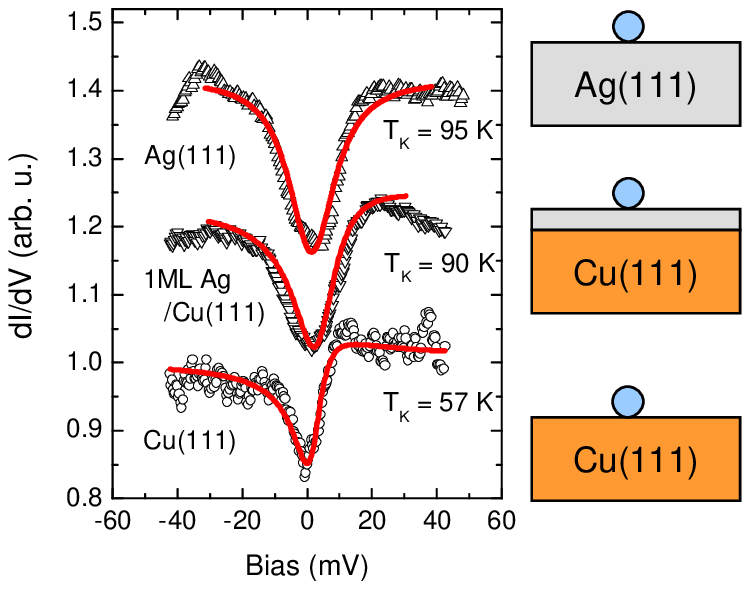}
 \caption[]{\label{fig1}
 Comparison of experimental $dI/dV$ spectra taken with the tip above a single Co adatom on a 1ML Ag film on Cu(111) with
 that on the clean Cu(111) and Ag(111) surfaces. Spectra are fitted to Eq. \ref{FanoEq} with $T_K$ as indicated,
 average values of sets of measurements are
 given in table~\ref{table1}. For 1ML of Ag the width of the
 resonance which is proportional to $T_K$ is already that of the
 adatom on Ag(111).
 }
\end{figure}

\begin{figure}
\includegraphics{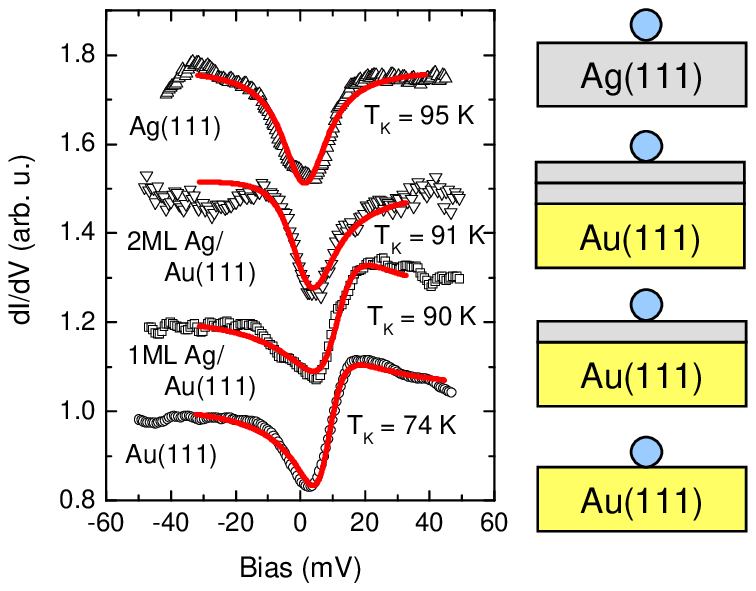}
 \caption[]{\label{fig2}
Same as Fig~\ref{fig1} but for a Co adatom on 1 and 2 ML of Ag on
Au(111) compared to Au(111) and Ag(111). Already for 1ML of Ag the
resonance width is the same as that of the adatom on Au(111).
 }
\end{figure}

In Fig.~\ref{fig1} we compare the experimental $dI/dV$ spectra
taken with the STM-tip on top of a Co adatom on 1 monolayer (ML)
of Ag on Cu(111) with those of a Co adatom on the clean (111)
surfaces of Cu and Ag. By preparing islands of Ag on Cu(111) we
were able to probe  Co on Cu(111) and Co on Ag/Cu(111) with the
same microscopic STM tip. Therefore a broadening of the Ag/Cu(111)
spectra due to experimental artifacts can be excluded. (The Co on
Ag(111) spectrum was taken with a different tip on a Ag(111)
crystal.) The spectra were fitted to the line shape given by the
Fano expression:
\begin{equation}
\frac{dI}{dV} \propto \frac{(q+\epsilon)^2}{1+\epsilon^2}
\label{FanoEq}
\end{equation}
where $\epsilon = \frac{eV-\epsilon_K}{\Gamma}$ is the normalized
energy and $\epsilon_k$ is the position of the resonance of width
$\Gamma$ relative to the Fermi energy.\cite{Ujsaghy00, Plihal01}
In the following we will identify the Kondo temperature $T_K$ with
the half width $\Gamma / k_B$ of the measured resonance. We find
that for a Co adatom on one ML of Ag on Cu(111) $T_K$ is within
the error bounds the same as that of a Co adatom on Ag(111). A
systematic study of the influence of the known reconstruction of
the Ag film on Cu(111) was not yet undertaken. We expect, however,
that it will be similar to the situation encountered at Au(111).
Here an influence of the surface reconstruction on the line shape
but not on $T_K$ was observed.\cite{Madhavan01}

The spectra of single Co adatoms on 1 or 2 ML of Ag on Au(111) are
shown in Fig.~\ref{fig2}. They yield the same finding as the
Ag/Cu(111) case. In the Ag/Au case it becomes more obvious that
the line shape (parameter $q$) is a function of Ag layer
thickness. The line shape is related to the electronic structure
of the host and the balance between tunneling into adsorbate and
into host electronic states
respectively.\cite{Madhavan98,Li98,Ujsaghy00,Plihal01} Since $q$
of Co on the Ag overlayer systems deviates from that of Co on
Ag(111) whereas $T_K$ does not, this parameter apparently reflects
single-particle electron wave functions and scattering processes
rather than the many-electron properties of the systems which we
focus on in this paper. What can already be seen is that
electronic properties on length scales much shorter than the
spin-correlation length (expected to be of the order 100~nm in the
systems studied here) determine $T_K$.

In table~\ref{table1} we summarize the average $T_K$ of the
resonance of Co adatoms on the various surfaces that are of
concern here. We also quote the Fano-parameter $q$, parameters of
the surface state at $\overline{\Gamma}$ like the onset energy
$E_0$, and effective mass $m^*$, as well as the band edge
positions of the sp-bands at the L-point of the bulk Brillouin
zone $E_{L_{2'}}$ and $E_{L_{1}}$, and the work function $\phi$.
The purpose of the latter parameters is to discuss in the
following whether they play a decisive role in determining the
measured Kondo temperatures of Co adatoms.

\begin{table*}
\caption{\label{table1}Average Kondo temperature $T_K$ and line shape
parameter $q$ for Co adatoms on noble metal surfaces and monolayer
systems. Also given are the surface state onset $E_0$ and its
effective mass $m^*$, the bulk band edge energies at $L$
$E_{L_{2'}}$ and $E_{L_1}$, and the work-function $\phi$. For the
overlayer systems $E_{L_{2'}}$ and $E_{L_1}$ were taken to be
those of the underlying bulk crystal, whereas $\phi$ was assumed
to be that of Ag. All energies are given relative to the Fermi
energy. ($^*$ denotes data measured in this paper)}
\begin{ruledtabular}
\begin{tabular}{llllllll}
substrate & $T_K$ & $q$ & $E_0$ & $m^*$ & $E_{L_{2'}}$ & $E_{L_{1}}$ & $\phi$ \\
Cu(111) & 54$\pm$2 K \cite{Manoharan00, Knorr02} & 0.2 & -0.44 eV \cite{Jeandupeux99}& 0.38$m_0$ \cite{Jeandupeux99}& -0.9 eV \cite{Chulkov99}& 4.25 eV \cite{Chulkov99}& 4.94 eV \cite{Chulkov99}\\
Au(111) & 76$\pm$8 K $^{*,}$\cite{Madhavan98} & 0.7 & -0.51 eV \cite{Buergi02} & 0.27$m_0$ \cite{Buergi02}& -1.0 eV \cite{Chulkov99}& 3.6 eV \cite{Chulkov99}& 5.55 eV \cite{Chulkov99}\\
Ag(111) &  92$\pm$6 K \cite{Schneider02}& 0.0 & -0.065 eV \cite{Jeandupeux99}& 0.40$m_0$ \cite{Jeandupeux99}& -0.4 eV \cite{Chulkov99}& 3.9 eV \cite{Chulkov99}& 4.56 eV \cite{Chulkov99}\\
1ML Ag/Au(111) & 88$\pm$10 K $^*$ & 0.8 & -0.27 eV \cite{unpublished}& 0.3 $m_0$ \cite{unpublished}& -1.0 eV & 3.6 eV & " \\
2ML Ag/Au(111) & 95$\pm$10 K $^*$& -0.1 & -0.2 eV \cite{unpublished} & 0.4 $m_0$ \cite{unpublished}& " & " & "\\
1ML Ag/Cu(111) & 92$\pm$10 K $^*$& 0.15 & -0.23 eV \cite{Wessendorf04} &  & -0.9 eV & 4.25 eV & "\\

\end{tabular}
\end{ruledtabular}
\end{table*}

$T_K$ of a magnetic impurity system depends very sensitively on
the details of the hybridization of the atomic levels of the
impurity with the host electronic system. In fact, from the Kondo
model it follows that

\begin{equation}
T_K \approx D e^{-\frac{1}{2 J \rho_F}} \label{KondoModelTK}
\end{equation}

where $D$ is the width of the substrate band, $J$ is the
antiferromagnetic ($J>0$) coupling of the impurity spin with the
spins of the substrate electrons and $\rho_F$ is the substrate's
density of states at the Fermi energy. $J$ depends on the coupling
matrix elements of the host electrons with the impurity. The
exponential dependence of $T_K$ on $J$ makes $T_K$ very sensitve
to details of adatom hybridization allowing to discuss the
contribution of different bands towards $J$.

First we want to discuss whether the properties of the
(111)-surface state determine $T_K$. If we look at the
surface-state band in these systems, we see that a change in $E_0$
or $m^*$ does not cause a systematic change in $T_K$. Although in
the overlayer systems the surface-state onset energy will shift
only within the first 5-10 Ag layers to the value of the Ag(111)
surface,\cite{Bendounan02,Wessendorf04} there occurs no change in
$T_K$ after the first layer. Furthermore, since $\rho_F =
\frac{m^*}{\pi \hbar^2}$  for a two-dimensional electron gas, the
surface-state properties would not only enter
Eq.~\ref{KondoModelTK} via the band width but also via $\rho_F$.
We note that systematic changes in $T_K$ due to the surface-state
band are not observed.

It was argued by Lin et al. \cite{Jones03} that the surface-state
electrons play a major role in the formation of the Kondo state
for Co on Cu(111) through their dominant contribution to $J$. Most
of that dominance is linked to the fact that surface-states are
naturally normalized to an area rather than a volume leaving the
decay of the state into the bulk crystal as a weight determining
parameter. To see whether the mere presence of a surface state
together with its decay determines $T_K$ we argue that in Ag(111)
the surface state is almost like a surface resonance with weight
up to the 10th crystal layer from the surface\cite{Chiang85}
whereas in Cu(111) the surface state decays much more rapidly.
Consequently, the weight of the surface state at the surface is
much smaller in Ag(111) than in Cu(111). By producing an overlayer
of 1ML Ag on Cu(111) the surface state still decays rapidly into
the bulk and one would therefore expect a similar contribution of
this state to $T_K$ as on the unmodified Cu(111) which appears not
to be the case.

The decay of the surface state into the bulk is related to
the positions of the bulk band edges at $L$. The band edges
determine the energetic positions of states with large momentum
perpendicular to the surface plane that could lead to differences
in the hybridization of the Co adatom. However, inspection of
table~\ref{table1} tells that this does not influence $T_K$: in
the overlayer systems the bulk electron states are unchanged and
so is the bandgap at $L$, but $T_K$ deviates greatly from that of
the impurity on the clean host crystal.

Finally, also the surface dipole, i.e. the work function, can be
shown not to influence $T_K$. If the decay length of wave
functions into the vacuum and the adsorbate position relative to
the surface layer determined $T_K$ this would explain why $T_K$ on
the overlayer systems is that of Co on Ag(111). However, Co on
Cu(111) would have the highest degree of hybridization due to its
short binding length (as estimated from a hard sphere model)
followed by Ag(111) and then by Au(111) due to
$\phi^{Ag}<\phi^{Au}$. This is in contradiction to
$T_K^{Au}<T_K^{Ag}$. From the data presented here we can therefore
conclude that changes in the sp-bandstructure including the
surface states on the noble metal (111) surfaces do not induce a
change in the Kondo temperature of magnetic adatom systems. This
leads us to the conclusion that the interaction determining the
Kondo temperature of a surface impurity system is very local in
nature, probably involving the d-orbitals of the substrate atoms.

\begin{figure}
\includegraphics{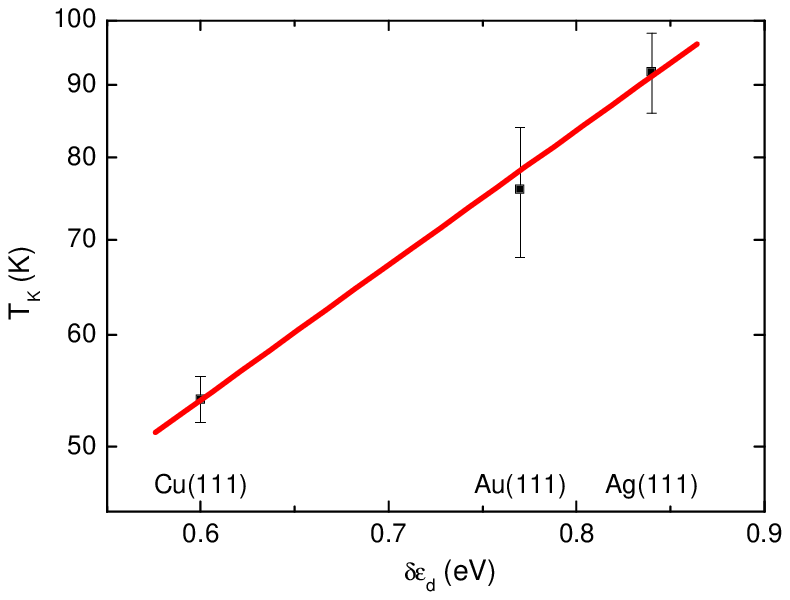}
 \caption[]{\label{fig3}
 Logarithmic plot of the Kondo temperature of single Co adatoms on Cu, Au, and Au(111)
 plotted vs. the d-level shift $\delta\epsilon_d$ calculated with
 LMTO-ASA\cite{Ruban97} for Co in the first crystal layer.
}
\end{figure}

A key to understanding the trend of Kondo temperatures of single
magnetic surface impurities could lie in studying connections
between calculated single particle properties and their mapping
onto the many-body problem. We would like to discuss such an
example in the following to trigger more theoretical
investigations. To do so, we have to use the expression for the
Kondo temperature in the more general single impurity Anderson
model instead of the Kondo model (Eq.~\ref{KondoModelTK}). In the
Anderson model the coupling constant $J$ is replaced by
(Schrieffer-Wolff transformation \cite{Schrieffer66})
\begin{equation}
\frac{1}{J} = \frac{\pi\rho_{F}}{\Delta U}|\epsilon_d|\cdot
|\epsilon_d + U| \label{SIAM_J}
\end{equation}
where $\epsilon_d$ is the energy of the impurity level, $\Delta$
its width acquired through hybridization, and $U$ is the on site
Coulomb repulsion. When comparing $T_K$ values of a specific
impurity (in our case Co) on different host metals a strong
variation in $T_K$ is expected from changes in $\epsilon_d$ as
this can vary by hundreds of meV. In Fig.~\ref{fig3} we
demonstrate that there exists a linear relation between the
calculated $\delta\epsilon_d$ for Co as a first layer impurity
\cite{Ruban97} and the logarithm of the measured $T_K$ for the three
systems Cu, Au and Ag(111). The calculated $\delta\epsilon_d$ were
originally used in conjunction with the model by Hammer and
N{\o}rskov to understand trends in the reactivity of transition
and noble metal surfaces.\cite{Hammer96}

The observed relation between $\delta\epsilon_d$ and $T_K$
indicates that the trend of the Anderson model $\epsilon_d$ for
the three surfaces can be written as $\epsilon_d = \epsilon_{d0} +
\delta\epsilon_d$ where $\epsilon_{d0}$, and $\Delta$ and $U$ do
not vary between the three systems or are a property of the
adatom. We note that the calculation of Ref.~\onlinecite{Ruban97}
predicts the lowest degree of hybridization for Co on Ag(111) and
the highest for Cu(111) consistent with our earlier observation of
a simple scaling law for $T_K$ of Co on other noble metal
surfaces.\cite{Wahl04}

A similar relation can be found between $T_K$ and the calculated
orbital moments of Co in Ag, Au, and Cu.\cite{Frota-Pessoa04} An
interesting aspect of this is that it might lead to an
understanding of the interplay between Kondo physics, spin-orbit
coupling and the magnetic anisotropy energy.\cite{Ujsaghy98} At
surfaces the anisotropy energy of single impurities can reach
values comparable to the Kondo temperatures discussed
here.\cite{Gambardella03}

In conclusion, we have determined the Kondo temperature $T_K$ of
single Co adatoms in interaction with Ag mono- and bilayers on
Cu(111) and Au(111) by STS and compare the results to $T_K$ found
for Co on the (111) surfaces of Cu, Ag and Au. We observe that
$T_K$ on the Ag overlayers is that of the Co adatom on Ag(111)
already for the first monolayer of Ag on either Cu or Au(111)
despite the fact that $T_K^{Cu} < T_K^{Au} < T_K^{Ag}$ for the
homogenous substrates. We find that properties of the surface
states and the energetic positions of the sp-derived bulk bands
found in these substrates cannot explain the trend in $T_K$. Since
the topmost crystal layer determines $T_K$ it is suggested that a
more local interaction of the Co d-levels with the d-bands of the
host metal is decisive. To further explore the Kondo properties of
Co on the noble metal (111) surface a correlation of the trend of
$T_K$ with the trend of the calculated shift of the impurity's
d-level on the different surfaces is shown. These single-electron
calculations have been used to explain the reactivity of 3-d
impurity sites at surfaces. The link between chemistry at the
surface and the many-body Kondo effect is established by the fact
that both properties depend sensitively on the details of the
impurity-host interaction. This observation might be useful to
gain a deeper understanding of the Kondo physics of adatom
systems.

\end{document}